\documentclass[conference]{IEEEtran}
\IEEEoverridecommandlockouts
\usepackage{cite}
\usepackage{amsmath,amssymb,amsfonts}
\usepackage{graphicx}
\usepackage{textcomp}
\usepackage{xcolor}
\usepackage{amsthm}
\usepackage{algorithm}
\usepackage{algpseudocode}
\algdef{SE}[DOWHILE]{Do}{doWhile}{\algorithmicdo}[1]{\algorithmicwhile\ #1}%

\usepackage{float}
\usepackage{subcaption}

\def\BibTeX{{\rm B\kern-.05em{\sc i\kern-.025em b}\kern-.08em
    T\kern-.1667em\lower.7ex\hbox{E}\kern-.125emX}}
\newtheorem{theorem}{Theorem}

\begin{document}

\title{A Line Graph-Based Framework for Identifying Optimal Routing Paths in Decentralized Exchanges
}

\author{\IEEEauthorblockN{Yu Zhang}
\IEEEauthorblockA{\textit{BDLT, IfI Department} \\
\textit{University of Zurich,}\\
Zurich, Switzerland 
\\Email: zhangyu@ifi.uzh.ch}
\and
\IEEEauthorblockN{Yafei Li}
\IEEEauthorblockA{\textit{Agroecology and Environment,} \\
\textit{Agroscope}\\
Zurich, Switzerland}
\and
\IEEEauthorblockN{Claudio Tessone}
\IEEEauthorblockA{\textit{BDLT, IfI Department} \\
\textit{University of Zurich,}\\
Zurich, Switzerland}
}

\maketitle

\begin{abstract}
Decentralized exchanges (DEXs), such as the CPMM-based (constant product market maker) Uniswap V2, play a pivotal role in the blockchain ecosystem by facilitating trading between any pairs of tokens. Trading between a pair of tokens involves providing a fixed amount of one token to receive the maximum possible amount of another token.
Despite active trading activities in DEXs, academic research on identifying optimal trading paths remains limited. This paper introduces a novel approach called the `line-graph-based algorithm', designed to efficiently detect the optimal trading paths for any token pair within a DEX.
After comparing the line-graph-based algorithm to the Depth-First-Search (DFS) algorithm commonly used by several famous DEXs in routing detection under the linear routing scenario, such as Uniswap, SushiSwap, PancakeSwap, QuickSwap, and BakerySwap, we found that the line-graph-based algorithm can help traders detect more profitable or at least equally profitable paths than DFS in all cases and the gas fee cost by the line-graph-based algorithm is comparable to that by the DFS algorithm after testing on a token graph with around 100 tokens and 200 liquidity pools from Uniswap V2 on October 30, 2022 and 2023, respectively.
For example, by selling tokens worth 10,000 dollars, our routing algorithm outperforms the DFS algorithm by more than 50\% in around 40\% cases, and both algorithms have the same performance in the remaining 60\% cases with each case meaning a token pair trading.
The computation complexity of the line-graph-based algorithm is $O(|E|\cdot(\sum {d_i}^2-2|E|))$, which is higher than DFS's computation complexity $O(|V|+|E|)$. $|V|$, $|E|$, and $d_i$ denote the number of nodes (token), number of edges (liquidity pool), and the degree of $i^{th}$ node in the token exchange graph, respectively. 
The average running time of the line-graph-based algorithm with Python is about 0.1 seconds and can be faster if the code is deployed with Rust, like 0.01 seconds.
These results show that this novel algorithm has a large potential application in industry.
\end{abstract}


\section{Introduction}
Decentralized finance (DeFi) has emerged as a prominent field within the blockchain ecosystem, leveraging blockchain technology to create financial applications that are open, transparent, and accessible to anyone. A crucial field of DeFi is the decentralized exchange (DEX) which allows peer-to-peer trading of tokens without the necessity of intermediaries with the mechanism of AMM (automatic market maker). In a DEX, there are plenty of trading pairs, called liquidity pools with tokens deposited pro-rata by liquidity providers. Traders can sell one kind of token into the liquidity pool to receive another kind of token with the exchange rate determined by the predefined AMM function. The main function types of AMMs include Constant Product Market Maker (CPMM), Constant Sum Market Maker (CSMM), Constant Mean Market Maker (CMMM), et al.. The commonly and popularly used AMM rule is CPMM. It relies on a constant product formula to determine the exchange price which states that the product of the quantities of two assets in a liquidity pool remains constant and has been popularly applied in many large DEXs such as Uniswap, Pancakeswap, SushiSwap, Balancer, et al..

In DEXs, many arbitrage opportunities exist because of price disparity. For example, research found that the cyclic arbitrage could be more than one million dollars sometime before 2022 in Uniswap V2. There is already research that focuses on identifying arbitrage opportunities on DEXs \cite{zhang2024improved,10426894} by combining the line graph of the DEX token exchange graph with the Bellman-Ford-Moore algorithm \cite{zhou2021just} or by the depth-first-searching algorithm \cite{yan2024optimizing}.

However, regular trading between token pairs should and must remain the primary activity in DEXs, as these platforms are fundamentally designed for token exchange rather than arbitrage. Unlike arbitrageurs, who primarily monitor markets to identify price disparities and exploit cyclic arbitrage opportunities using flash loans by optimizing token input amounts, regular traders typically provide a fixed amount of a specific token, aiming to receive the maximum possible quantity of the desired token. Flash loan is a unique characteristic of DeFi compared to traditional finance \cite{ruetschi2024decentralized}, which facilitates cyclic arbitrages.
For the sake of simplicity and clarity in description, we would call the sold token in trading as source token and the received token as target token.
The distinction between arbitrage and trading highlights that the challenge of detecting optimal trading paths for traders is fundamentally different from that of arbitrage detection. In the following parts, the term ``trading" or ``exchange" always refers to buying one token as much as possible by selling a fixed amount of another token. 
% The problem of detecting the optimal trading path between a pair of tokens with a fixed amount of input is simple in logic. 
When exchanging a specific quantity of one token for another in a DEX with a small token exchange graph (constructed from tokens and liquidity pools), it is feasible to enumerate all possible trading paths and select the one yielding the maximum output. However, as the size of the token exchange graph grows, the number of potential trading paths increases exponentially, rendering enumeration methods impractical in real-world scenarios.
For instance, in a DEX with hundreds of liquidity pools, millions of possible paths may exist for exchanging a single pair of tokens.

% what mechanism do they use for trading?
In industry, relevant DEXs apply very basic algorithms to enumerate a few possible trading paths between the traded pairs of tokens.
Uniswap V2 and Pancake use the Depth First Search (DFS) to help traders search for optimal trading paths. DFS is not a good choice for optimal trading because this algorithm is not designed for optimal path detection. For example, a large proportion of possible paths will be ignored which makes it hard to get a good trading path. The running results of the algorithm are very random. 

Currently, there is little academic research addressing the problem of optimal trading path detection. In this paper, we propose a novel line-graph-based algorithm to assist traders in identifying more profitable trading paths compared to the traditional DFS algorithm. 
An improved routing algorithm not only benefits traders but also enhances the market efficiency of a DEX. Market efficiency, in this context, refers to fewer arbitrage opportunities within the DEX.
We will also compare the profitability performance of our algorithm and the DFS algorithm on Uniswap V2 data to show how much our algorithm outperforms the DFS algorithm.

The paper is organized as follows: the second part is related work about routing algorithms in DEXs; in the third part, the data that will be used to compare different algorithms' performances is described; in the fourth part, we put forward the newly developed method for trading in DEXs; in the fifth part, we compare our method to the commonly used methods in industry, the DFS algorithm, from different perspectives, then we discuss our method a bit and conclude the paper.

\section{Related Work} 

Trade routing is an indispensable part of DEXs. Different DEX platforms may use different algorithms to detect optimal trading paths for traders. For example, Uniswap, PancakeSwap, SushiSwap, et al. use DFS-based algorithms. The DEX aggregator 1inch used a similar algorithm but upgraded to another algorithm that has not been published. The DFS-based algorithms are efficient in the calculation but can not always obtain more profitable trading paths for traders.

However, there is limited research in the academic field focusing on routing in DEXs. \cite{danos2021global} promoted a convex-optimization-based method to solve arbitrage and trade routing problems in DEXs. Then, \cite{angeris2022optimal} systematically introduced the optimization-based method to solve the arbitrage maximization problem in DEXs, which can be extended to solve the trade routing problem. The convex-optimization-based method can provide the optimal trading path in theory but the calculation of convex optimization is very complex and unreliable when the token graph is large, which makes it unable to be applied in real calculations. For example, the trade may need to use most liquidity pools if we calculate it by convex optimization methods, which is shown in the example in Appendix \ref{problem_convex_optimization}.

In this paper, we will provide a DEX routing algorithm that is more profitable than DFS-based algorithms and has the potential to be deployed in DEXs such as Uniswap DEX.

\section{Data Description}
\label{data_description}
There have been around four hundred thousand tokens as of now in Uniswap DEX. Token liquidity pool data can be used to construct the token graph where nodes present tokens and an edge is created between two tokens if there is a liquidity pool consisting of the two tokens. In this way, all token pools can be connected as a token graph $G(V, E, R)$ with $V$ as the token set, $E$ as the token liquidity pool set, and $R$ as the set of token reserves in corresponding liquidity pools (edges) which is in the form of ${\{X_1:x_1, Y_1:y_1\}, \{X_2:x_2, Y_2:y_2\}, \cdots}$. The capital letters denote tokens, and the lower-case letters denote the reserve of the specific tokens in corresponding liquidity pools.

For this research, we select only those active and large liquidity pools with similar filtering processes as in \cite{zhang2024improved}, the liquidity pool filtering rules is described in Appendix \ref{poolfilter}.
After filtering, the graph has ninety-four tokens and one-hundred-and-ninety-two liquidity pools with the most liquidity. 

For Uniswap V2, the Constant Product Market Maker (CPMM) is used to determine the token exchange and the marginal output decreases with more input. A detailed description of CPMM function of Uniswap V2 can be found in \cite{zhang2024improved,xu2023sok, gogol2024sok} and Appendix \ref{amm}.

\section{Method}

The token graph constructing procedures and the line-graph-based algorithm were described step by step in this part, which contains:
\begin{enumerate}

    \item Firstly, the tokens' liquidity pool data is used to construct the token graph $G(V, E, R)$ with $V$ being the token set, with $E$ being the unidirectional tradable token pairs, and with $R$ being the set of token reserves in corresponding liquidity pools (edges) which is in the form of $\{(v_i,v_j):(r_i,r_j), \cdots\}$, $i\neq j$. We can also write the token graph $G(V, E, R)$ as $G$ for short for simplicity in later parts if necessary.
    \item Secondly, the line graph $L(G)$ is constructed according to the token graph $G(V, E, R)$.

    \item Thirdly, we describe the line-graph-based algorithm for detecting the optimal trading path in DEXs under the scenario of linear routing. 
    % We will also prove that the detected path is almost optimal for traders in theory and provide the computation complexity of the method. 
\end{enumerate}

\subsection{Constructing Token Graph}\label{construct_token_graph}

The token graph $G(V,E,R)$ is directed, where $v_i \in V$ denotes the $i^{th}$ token in the token set $V$ ($i=1,2,...N$), $e_{ij} \in E$ denotes the directed edge from token $v_i$ to $v_j$ and is denoted as $(v_i,v_j)$. $(v_i,v_j):(r_i,r_j) \in R$ denotes the token reserves of the token tuple $(v_i,v_j)$ for which the order matters, namely the reserve of $v_i$ is $r_i$ and the reserve of $v_j$ is $r_j$ in the edge $(v_i,v_j)$.
$(v_i,v_j)\neq (v_j,v_i)$ in our settings.

The edge $e_{ij}$ represents the unidirectional exchange by providing token $v_i$ to obtain $v_j$. It can be expressed as $e_{ij}=(v_i, v_j)$ just as described above. For a liquidity pool containing a pair of tokens $v_i$ and $v_j$, there exist two different edges, the edge from token $v_i$ to $v_j$ ($e_{ij}$) and the edge from token $v_j$ to $v_i$ ($e_{ji}$), so, $e_{ij}\neq e_{ji}$. 
The edge $e_{ij}$ means that traders can only trade $v_i$ for $v_j$. In this paper, we will call it a \textbf{directed liquidity pool} which can also be expressed as $(v_i,v_j)$. A liquidity pool containing both $v_i$ and $v_j$ will be denoted as a set $\{v_i,v_j\}$, which means that traders can trade token $v_i$ for $v_j$ and vice versa. 

In $G(V,E,R)$, each edge $e_{ij}$ stores the reserve information of token $v_i$ and $v_j$ in the directed liquidity pool $(v_i,v_j)$ as a tuple $(r_i,r_j)$. In edge $e_{ji}$, the stored reserve information will be $(r_j,r_i)$. This information is critical for calculating the quantity of token $v_j$ that traders can get by inputting a specific quantity of token $v_i$.

In DEX, the token exchange is defined by the AMM rule which is denoted as $f$. $f$ defines the quantity relationship between one token's input and another token's output, and $\Delta r_j = f_{(r_i,r_j)}(\Delta r_i)$. Where $\Delta r_i$ denotes the input quantity of token $v_i$; $\Delta r_j$ denotes output quantity of token $v_j$; $(r_i,r_j)$, the reserve tuple of token $v_i$ and $v_j$, are parameters of $f$ and $f_{(r_i,r_j)}$ are different from $f_{(r_j,r_i)}$. The function $f$ has two properties: $\textbf{(1) it is monotonically increasing; (2) it is concave}$. Almost all market maker functions fulfill these two requirements, which means that our method applies to all kinds of DEXs in the DeFi system.

\subsection{Constructing Line Graph for Token Graph}\label{construct_line_graph}

Given the underlying graph $G(V, E, R)$, its line graph $L(G)$ can be constructed by the following steps:
\begin{enumerate}
    \item Constructing the vertexes in the line graph. The edges ($e_{ij}$) in the original underlying graph $G(V, E, R)$ are taken as the new vertices of the line Graph $L(G)$. A new vertex in the line graph $L(G)$ will be denoted as a tuple with its two entries representing the two tokens at the two ends of each edge ($e_{ij}$) in $G(V, E, R)$. For example, in $G(V, E, R)$, the edge $e_{ij}$ is a directed edge from $v_i$ to $v_j$. In $L(G)$, the corresponding new vertex will be $(v_i, v_j)$, $i, j = 1,2,...N $ and $i\neq j$. We also store tokens' reserve information $(r_i,r_j)$ belonging to the directed liquidity pool in the new vertex $(v_i,v_j)$ and it is denoted as $(v_i,v_j)=(r_i,r_j)$.
    
    \item Creating links for vertexes in the line graph. Any pair of new vertices in line graph $L(G)$ are linked if the last token of a new vertex is the first token of another new vertex. For example, assuming one new vertex in $L(G)$ is $(v_i, v_j)$ and another new vertex is $(v_j,v_l)$, then we add a link from the new vertex $(v_i, v_j)$ to vertex $(v_j,v_l)$ in $L(G)$. The new link from $(v_i, v_j)$ to $(v_j,v_l)$ in $L(G)$ can be denoted as $(v_i, v_j, v_l)$ or $(v_i, v_j)\rightarrow (v_j, v_l)$.

    % \item Store token reserve information to the corresponding edges of the line graph. 
    % The token reserve information of edge $(v_i, v_j, v_l)$ (or $(v_i, v_j)\rightarrow (v_j, v_l)$) in $L(G)$ is set as $(r_j, r_l)$ which is the reserve information of the edge from token $v_j$ to $v_l$ in $G(V, E, R)$.
    
     \item Cutting some unnecessary links for the line graph. For any pair of new vertices in $L(G)$, $(v_i, v_j)$ and $(v_k,v_l)$, if $j = k$, $i=l$ and $i \neq j$, then we cut both the two mutual links between the two vertices. This means we cut both the edge $(v_i, v_j, v_i)$ and edge $(v_j, v_i, v_j)$ in $L(G)$. $(v_i, v_j, v_i)$ represents the case that traders trade $v_j$ with $v_i$ and then trade $v_j$ back for $v_i$ in the same liquidity pool that contains both $v_i$ and $v_j$. It is impossible to make a profit from this kind of trade in theory and practice because of transaction fees and transaction tax paid to liquidity providers. So, we cut these links. 
     By cutting these links, the computation complexity is lower.
\end{enumerate}

The number of vertices in the new line graph ($M_{L(G)}$) equals the number of edges ($E_G$) in the underlying graph ($G$), namely, $M_{L(G)}=E_G$. The number of links in the line graph ($E_{L(G)}$) equals the sum of the degree's square of each node minus two times the number of edges in graph $G$, namely $E_{L(G)} =\sum {d_i}^2-2E_G$, where $d_i$ denotes the degree of token (node) $i$ in the underlying graph $G$ and $E_G$ denotes the number of edges of the underlying graph $G$.

To simplify the calculation again, we add an extra vertex for the line graph $L(G)$ with the following steps:

 \begin{enumerate}
     \item Construct the line graph $L(G)$ by steps as shown above.
     \item We assume that $v_1$ is the source token that we will consider and $(v_1, v_{n_1})$, $(v_1, v_{n_2})$, $...$, $(v_1, v_{n_k})$ are its neighbour link vertices in $L(G)$. Now, we add an extra node called $(O,v_1)$ to $L(G)$ and link this vertex to all the neighbor link vertices of $v_1$, namely, $(O,v_1)\rightarrow(v_1, v_{n_2})$, $...$, $(O,v_1)\rightarrow(v_1, v_{n_k})$. $(O,v_1)$ is also called the source vertex in $L(G)$, and $O$ is just a sign that is not important and can be any other sign.
     % \item 
     % The token reserve information of each extra added edge in $L(G)$ is set as that of the edge from the source token ($v_1$) to all tokens in the neighbor token set of $v_1$ in $G(V, E, R)$, respectively. For example, the reserve of the edge $(O,v_1)\rightarrow(v_1, v_{n_k})$ is $(r_1,r_{n_k})$.
 \end{enumerate}

\begin{figure}[h]
    \centering
    \includegraphics[width=0.5\linewidth]{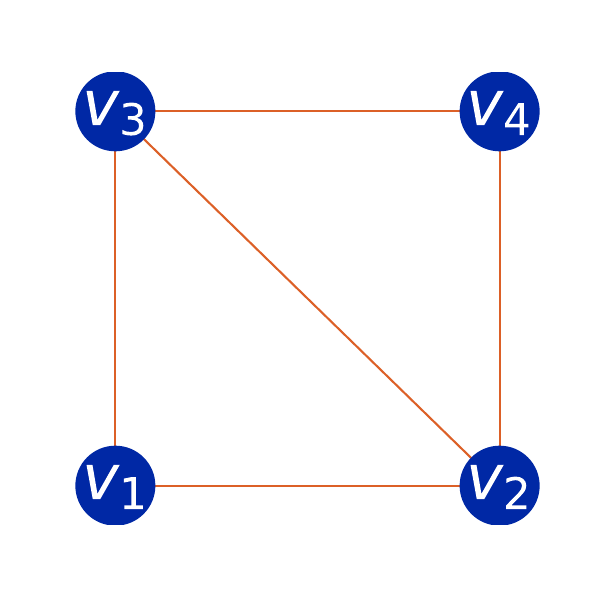}
    \caption{Orginal graph $G$ with 4  tokens and 5 liquidity pools. So, $M_G=4$ and $E_G=5$, where $M$ and $E$ denote the number of nodes and edges.}
    \label{origraph}
\end{figure}

\begin{figure}[h]
    \centering
    \includegraphics[width=1\linewidth]{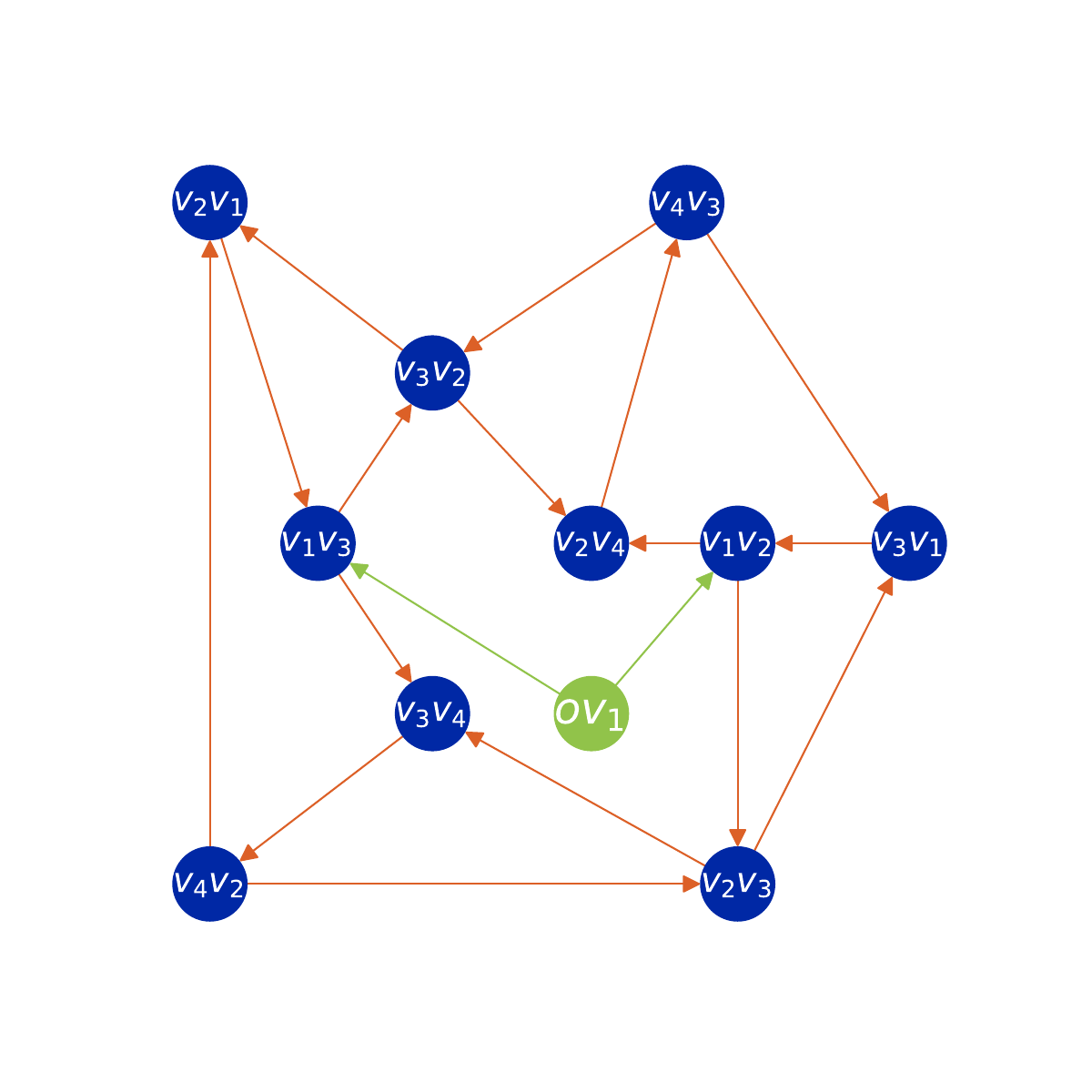}
    \caption{The line graph ($L(G)$) of the original graph $G$ with 4 tokens and 5 liquidity pools as shown in Fig.\ref{origraph}. The number of vertices in the line graph equals the number of edges in the original graph, namely, $M_{L(G)}=E_G$. The number of links in the line graph equals the sum of the degree's square of each token minus two times the number of edges in graph $G$, namely $E_{L(G)} =\sum {d_i}^2-2E_G$, where $d_i$ denotes the degree of token $i$ in the original graph $G$. The added extra node to simplify calculation is the green vertex $(O,v_1)$ which connects to the other two vertices, $(v_1,v_3)$ and $(v_1,v_2)$.}
    \label{linegraph}
\end{figure}

The above steps maybe a bit abstract. So, we take a simple example which may help readers understand the procedures better and more deeply.
Assuming an original token graph with 4 tokens ($v_1,v_2,v_3,v_4$) and 5 liquidity pools ($\{v_1,v_2\}$, $\{v_1,v_3\}$, $\{v_2,v_3\}$, $\{v_2,v_4\}$, $\{v_3,v_4\}$) is shown in Fig. \ref{origraph}, its line graph with more vertices and links can be constructed as shown in Fig. \ref{linegraph} if we follow the above line graph constructing rules and rules to add an extra vertex.

\subsection{Line-graph-based Algorithm in DEXs}

Though there are more nodes and links in the line graph, the situation that only a unidirectional link exists between two linked nodes will make our calculation easier. The advantage of the line graph is that all token combinations in order are still reserved. For example, all ways from the token $v_1$ to $v_4$ in Fig. \ref{origraph} are still in their corresponding line graph \ref{linegraph}.

For any token loops in the original graph, there will be another two corresponding loops in its line graph. For example, there is a loop $v_1\rightarrow v_2\rightarrow v_3 \rightarrow v_1$ in Fig.\ref{origraph}, the two corresponding loops in the line graph Fig.\ref{linegraph} are 
$v_1 v_2 \rightarrow v_2 v_3 \rightarrow v_3 v_1$ and $v_1 v_3 \rightarrow v_3 v_2 \rightarrow v_2 v_1$.

The algorithm starts from the extra vertex ($(O,v_1)$) with inputting $\epsilon 1$ units of source token $v_1$.
Then each link in $L(G)$ is iterated to calculate the maximal output of another token for a given input. For a link $(O,v_1)\rightarrow(v_1,v_{n_k})$, the input of $v_1$ is $\epsilon1$, then we calculate the $v_{n_k}$'s output based on the defined AMM rule $f$ using the reserve parameters stored in $(v_1,v_{n_k})$. At last, we store the calculated output token $v_{n_k}$ to vertex $(v_1,v_{n_k})$. The iterations procedure is a bit similar to that in the Bellman-Ford-Moore method.

In arbitrage detection using the Bellman-Ford-Moore algorithm, each arbitrage opportunity corresponds to a negative loop where the cumulative exchange rate exceeds one. The algorithm gets trapped in these loops, continuously updating token distance values. In contrast, token trading differs fundamentally, even within arbitrage loops, preventing infinite iterations and ensuring the algorithm terminates in finite steps. This is formally established in the following theorem and proof.

\begin{theorem}
    In an arbitrage loop with the accumulative product of the token exchange rate larger than one, given the market maker function $f$ monotonically increasing and concave, for any fixed amount of input of a specific token in the arbitrage loop, the number of traversal round to get the higher output of any tokens than the previous round is finite.  
\end{theorem}

\begin{proof}
    Assuming an arbitrage loop $v_1\rightarrow v_2\rightarrow \cdots v_k\rightarrow v_1$ and the reserve of a liquidity pool ($\{v_i,v_j\}$, a set) containing token $v_i$,$v_j$ is $r_{{\{v_i,v_j\}}_i}$ and $r_{{\{v_i,v_j\}}_j}$ and $v_i\rightarrow v_j$ is part of the arbitrage loop.

    The traversal starts from the token $v_1$ along the arbitrage loop.
    We input a fixed amount $\epsilon (>0)$ of token $v_1$  in the arbitrage loop and output $\epsilon_j^1$ token $v_j$ in the first round. If we get $\epsilon_j^2$ token $v_j$ in the second round and $\epsilon_j^2>\epsilon_j^1$, then the traversal continues. Otherwise, the exchange stops.
    The total outputted token $v_j$ in the liquidity pool $\{v_i,v_j\}$ is $\sum_{m=1}^M{\epsilon_j^m}$ with $\epsilon_j^m$ non-decreasing with $m$. Because the total reserve of token $v_j$ from liquidity pool $\{v_i,v_j\}$ is finite, so, the number of exchange rounds to get a higher output token is finite.  
\end{proof}

This proof shows that the number of exchange steps is finite even in the case of an arbitrage loop. However, the result also applies to all other cases, which means that the calculation to get a trading path by iterative dynamic programming is finite.

Based on this theorem, we give the line-graph-based algorithm (Algorithm \ref{Algorithm 1}) to calculate the more profitable trading path by selling a fixed amount of one token for buying another target token.

\begin{algorithm}[H]
\caption{Linear Routing Algorithm in a DEX}   \label{Algorithm 1}
\textbf{Problem setting:} \textit{We need to buy the maximal units of token $v_N$ ($\Delta r_n$) with $\epsilon_1$ units of token $v_1$ in a DEX.}

    \hspace*{\algorithmicindent} \textbf{Input}: $\epsilon_1$ units of $v_1$\\
    \hspace*{\algorithmicindent} \textbf{Output}: The maximal units of $v_n$
    \begin{algorithmic}[1]

    \State $L(G)$ $\gets$ Line graph of $G$ with source vertex $(O,v_1)$ and token reserve information $(v_i,v_j)=(r_i,r_j)$ on each vertex (directed liquidity pool) from a $DEX$.

    \State $f_{(r_i,r_j)}(\cdot)$ $\gets$ A monotonically increasing and concave market maker function with parameters $(r_i,r_j)$ corresponding to tokens' latest reserve in liquidity pool $(v_i,v_j)$ during the updating.
    \algstore{myalg}

\end{algorithmic}
\end{algorithm}

\begin{algorithm}[H]                 
\begin{algorithmic}[0] 
    \algrestore{myalg}
    \State $D$ $\gets$The quantity of token output dictionary from the source vertex $(O,v_1)$ to all other vertices in $L(G)$. 
    Setting $D[(O,v_1)]=\epsilon_1$, $D[(v_i,v_j)]=0$ for $i,j \in (1,2,\cdots, N)$ and $i\neq j$.
    \State $P$ $\gets$Corresponding trading path dictionary. Setting $P[(O,v_1)] = [(O,v_1)]$, $P[(v_i,v_j)]=null$ for $i,j \in (1,2,\cdots, N)$, $i\neq j$.

    \State $R$ $\gets$ A dictionary, keep reserve information in each exchange step. 

\vspace{0.1cm}
    \Do
        \For{Each edge $(v_i, v_j)\rightarrow (v_j, v_l)  $ in $L(G)$}
            \If{$D[(v_i,v_j)]>0$}
                \If{$D[(v_j,v_l)]=0$}
                    \State $D[(v_j,v_l)]=f_{(r_j,r_l)}(D[(v_i,v_j)])$
                    \State $P[(v_j,v_l)]=P[(v_i,v_j)]+(v_j,v_l)$
                    \State $R[(v_j,v_l)]=R[(v_i,v_j)]+(r_j+D[(v_i,v_j)],r_l-D[(v_j,v_l)])$
                \Else
                    \If{$(v_j,v_l) \notin P[(v_i, v_j)]$)}
                        \State $d = f_{(r_j,r_l)}(D[(v_i,v_j)])$
                        \If{$d>D[(v_j,v_l)]$}
                            \State $D[(v_j,v_l)]=d$
                            \State $P[(v_j,v_l)]=P[(v_i,v_j)]+(v_j,v_l)$
                            \State $R[(v_j,v_l)]=R[(v_i,v_j)]+(r_j+D[(v_i,v_j)],r_l-D[(v_j,v_l)])$
                        \EndIf
                    \Else
                        \State $I = R[(v_i, v_j)].last\_index((v_j, v_l))$
                        \State $J = (v_l, v_j) \in R[(v_i, v_j)]? R[(v_i, v_j)].last\_index((v_l, v_j)): -1$
                        \If{$I > J$}
                            \State $d = f_{(R[(v_i, v_j)][I])}(D[(v_i,v_j)])$
                        \Else
                            \State $(r_j,r_l)= R[(v_i, v_j)][J][::-1]$ 
                            \State $d = f_{(r_j,r_l)}(D[(v_i,v_j)])$
                        \EndIf
                        \If{$d>D[(v_j,v_l)]$}
                            \State $D[(v_j,v_l)]=d$
                            \State $P[(v_j,v_l)]=P[(v_i,v_j)]+(v_j,v_l)$
                            \State $R[(v_j,v_l)]=R[(v_i,v_j)]+(r_j+D[(v_i,v_j)],r_l-D[(v_j,v_l)])$
                        \EndIf
                    \EndIf
                \EndIf
            \EndIf
        \EndFor
    \doWhile{D is updating}
    \vspace{0.1cm}

    \State $T$: The maximal amount of $v_n$.
    \State $TP$: The optimal trading path. 
    \begin{enumerate}
        \item $T = argmax_{v_k}(D[(v_k,v_n)])$  with $v_k \in N_{v_n}$,
    where $N_{v_n}$ denotes $v_n$'s neighbour token set.
        \item $I= where(D[(v_k,v_n)]=T)$, $where$ is a function to return the index of the item corresponding to $T$ in $D[(\cdot, v_n)]$. 
        \item $TP = P[I]$
    \end{enumerate}

    \Return $T$, $TP$
    \end{algorithmic}
    
\end{algorithm}

To prove that the algorithm can detect the optimal trading path, we put forward the following theorem which is a very strong condition:

\begin{theorem}
    In the line graph $L(G)$, if the paths from the source vertex to all other vertices are still being updated, routing paths with a higher amount of target tokens can be obtained.
\end{theorem}

\begin{proof}
    This theorem is apparent. 
    % If we assume that the optimal trading path has not been gotten yet, then we need to prove that the path will update. 
    
    Assuming a DEX owns three token liquidity pools, including a pool containing $v_1$ and $v_2$, a pool containing $v_1$ and $v_3$, and a pool containing $v_3$ and $v_2$. If the algorithm has updated the exchange with $\epsilon_1$ token $v_1$ for $\Delta v_3$ token $v_3$, and the exchange with $\epsilon_1$ token $v_1$ for $\Delta v_2$ token $v_2$, but not yet update the exchange with token $v_2$ for $v_3$.
    Assuming that one can exchange $\Delta v_2$ token $v_2$ for $\Delta v_3^`$ token $v_3$ and $\Delta v_3^`>\Delta v_3$. According to the line-graph-based algorithm, it continues to update the exchange with $v_2$ for $v_3$ and more profitable route is obtained. 
    % which is a contradiction with the assumption that the paths don't update even if when the trading path is not optimal.
\end{proof}

In algorithm \ref{Algorithm 1}, the stop standard of the algorithm is that all paths don't update anymore. If there is no arbitrage loop, then the line-graph-based algorithm is very similar to the Bellman-Ford algorithm, and the computation complexity will be $O(M_{L(G)}\cdot E_{L(G)})$ where $M_{L(G)}$ is the number of vertex and $E_{L(G)}$ is the number of edges in the line graph $L(G)$. But in case there exist arbitrage loops, then an extra number of iterations $K$ may be necessary to traverse these arbitrage loops, where $K$ should not be a large value, and the maximal computation complexity is $O(M_{L(G)}\cdot E_{L(G)}+K)\simeq O(M_{L(G)}\cdot E_{L(G)})$. 

\section{Experiment and Result}

In this section, we compare the performance of the line-graph-based algorithm to the commonly used depth-first-search (DFS) algorithm using Uniswap V2 liquidity pool data on 2022.10.30 and 2023.10.30. The DFS Algorithm \ref{Algorithm 6} is described in Section \ref{appendix}.

As described in Section \ref{data_description}, the token graph data from Uniswap V2 that we will apply in our comparison experiment contains around 100 tokens and 200 liquidity pools. We will run both the line-graph-based algorithm and the DFS algorithm on all pairs of different tokens from our constructed token graph to compare their performance. The total number of different token pairs is $10,000 = 100\times100$ and $8742= 94 \times 93$, respectively. The routing algorithm performance is compared by inputting tokens with a fixed value, measured in USD (\$). For example, assuming trading the source token $A$ to the target token $B$, if token $A$'s price in a centralized exchange is $p_A$ and the amount of money input is $M$, then the number of source token input ($A$) in both algorithms is $\frac{M}{p_A}$.

In a specific experiment, both algorithms will be applied to all token pairs to calculate the routing path by selling the corresponding source tokens, which have the same values measured in USD. We take the experiment on 2023-10-30 as an example for illustration. The analysis on 2022-10-30 is similar to that on 2023-10-30.
\begin{figure}[H]
\begin{subfigure}{0.5\textwidth}
\centering
\includegraphics[width = 0.9\linewidth]{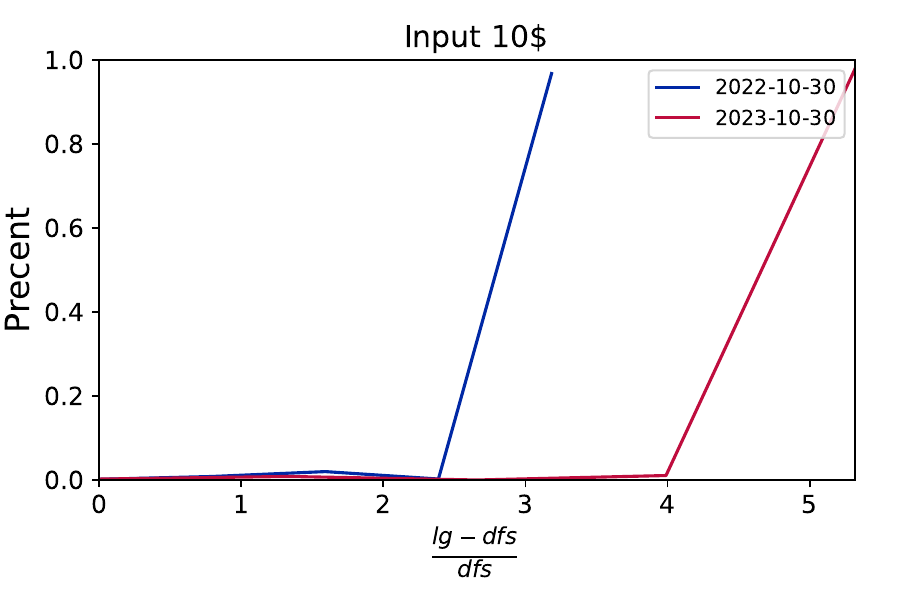}
% \caption{a}
\end{subfigure}  
\begin{subfigure}{0.5\textwidth}
\centering
\includegraphics[width = 0.9\linewidth]{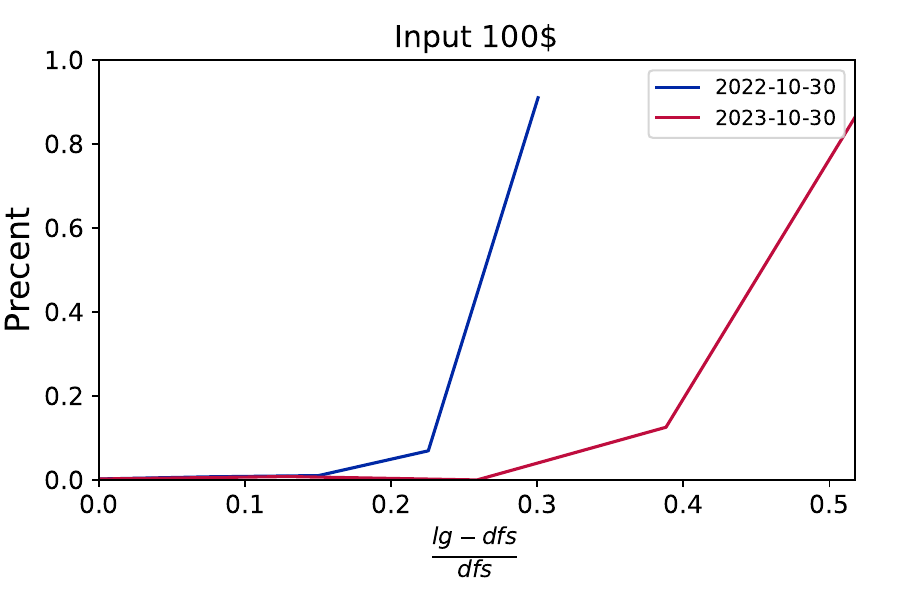}
% \caption{b}
\end{subfigure}   
\begin{subfigure}{0.5\textwidth}
\centering
\includegraphics[width = 0.9\linewidth]{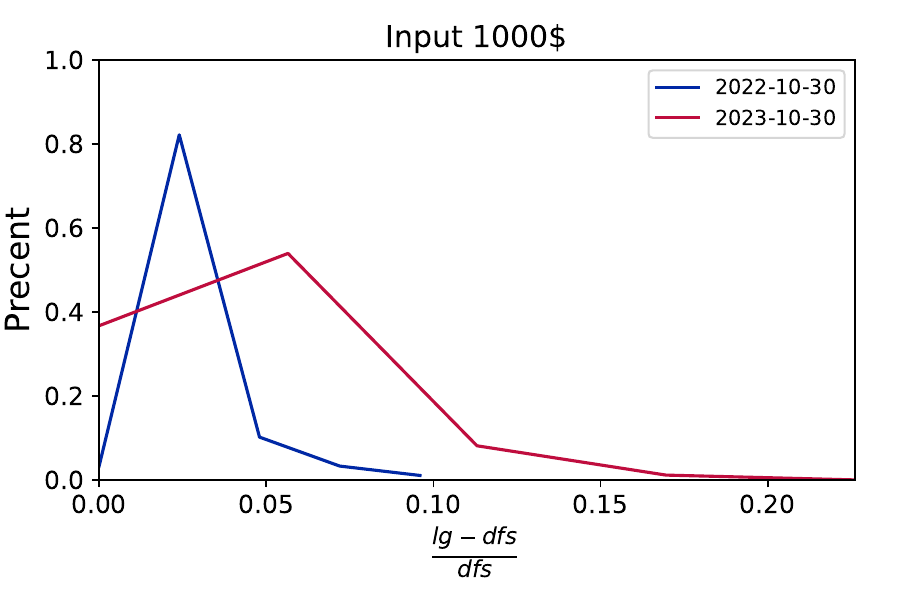}
% \caption{b}
\end{subfigure} 
\begin{subfigure}{0.5\textwidth}
\centering
\includegraphics[width = 0.9\linewidth]{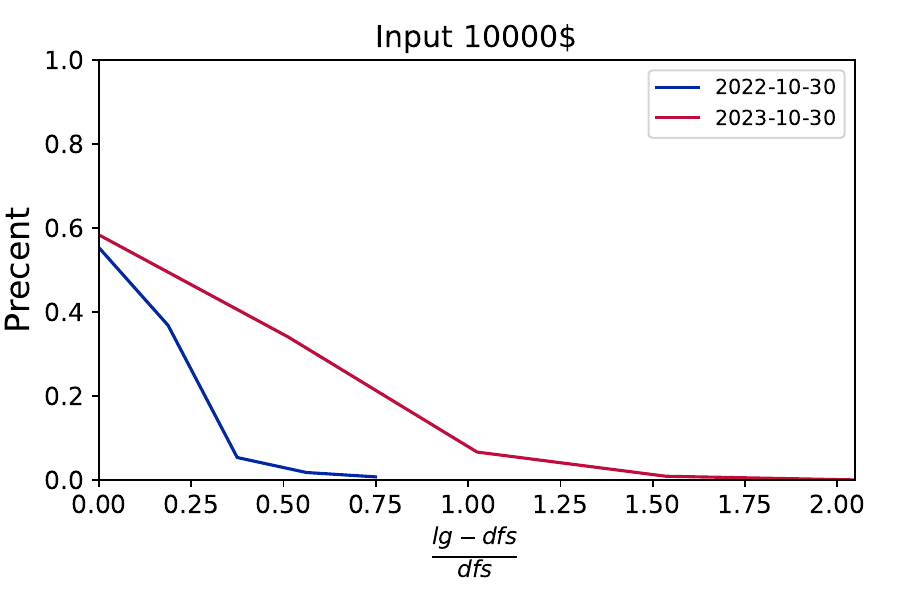}
% \caption{b}
\end{subfigure} 
\caption{The distribution of performance discrepancy ratio between the two algorithms, line-graph-based algorithm, and DFS algorithm, by inputting different amounts of money (10\$, 100\$, 1,000\$, and 10,000\$). $lg$ and $dfs$ represent the target token outputs for line-graph-based algorithm and DFS algorithm, respectively. The Y-axis shows the percentage of token pairs where the line-graph-based algorithm outperforms the DFS algorithm by the corresponding ratio.}
\label{algorithm_comp}
\end{figure}
Firstly, we set the amount of money input as 10\$, namely $M=10\$$, and compare the performance of two algorithms by running both of them on all 8742 pairs of tokens, whose result is shown in the first subfigure of Fig. \ref{algorithm_comp}. In this subfigure, $lg$ denotes the number of target tokens received by inputting a source token valued $M=10\$$ with the line-graph-based algorithm, and $dfs$ denotes the number of target tokens received with the same settings with the DFS algorithm. So, in this subfigure, the X-axis $\frac{lg-dfs}{dfs}$ measures the ratio to what extent the line-graph-based algorithm outperforms the DFS algorithm in profitability. The Y-axis measures the corresponding percent of token pairs where the line-graph-based algorithm exceeds the DFS algorithm by the specific ratio $\frac{lg-dfs}{dfs}$. The red line in the first subfigure of Fig.\ref{algorithm_comp} shows that the line-graph-based algorithm obtains an equal or higher number of target tokens compared to the DFS algorithm when running them in all 8742 pairs. In all token pairs, the number of received target tokens by running the line-graph-based algorithm is almost 5 times higher than that by running the DFS algorithm. 

In regular token trading, the number of source tokens input is a critical element that affects the evaluation of routing algorithms' performance. So, we repeat the same experiments by setting $M=100\$,1,000\$, 10,000\$$, and all these results are shown in the second to the fourth subfigure of Fig.\ref{algorithm_comp}.

When $M=100\$$, the line still has a monotonously increasing trend. In around 80\% of pairs, the line-graph-based algorithm outperforms the DFS algorithm 50\%, and the ratio is 40\% in the rest around 20\% of pairs.

When $M=1,000\$$ and $M=10,000\$$, the distribution lines of the performance discrepancy ratio between the two algorithms $\frac{lg-dfs}{dfs}$ have very different patterns compared to that in the experiments where $M=10\$$ and $M=100\$$ and are monotonously decreasing, which means that the line-graph-based method outperforms the DFS algorithm larger and larger in less and less proportion of token pairs when $M$ is fixed and large enough. 

However, if we compare the scale of the x-axis of the third and fourth subfigures in Fig.\ref{algorithm_comp} which corresponds to the case of $M=1,000\$$ and $M=10,000\$$, respectively, it shows that the extent of line-graph-based algorithm outperforming DFS algorithm is higher if the trade size is larger, namely larger $M$. This conclusion is contrary to the one obtained by comparing the first two subfigures in Fig.\ref{algorithm_comp}.

To compare the performance of the two algorithms systematically, we conduct more experiments by setting $M$ ranging from 100\$ to 10,000\$ with a step size of 100\$. As shown in the third and fourth subfigures in Fig.\ref{algorithm_comp}, the value of $\frac{lg-dfs}{dfs}$ is zero for a large proportion of token pairs when $M$ is large.
So, we first calculate the proportion of token pairs where $\frac{lg-dfs}{dfs}$ is larger than zero, which is a useful statistic to understand the two algorithms' performance. Though we have $\frac{lg-dfs}{dfs} \geq 0$ in all cases, there is a high proportion of token pairs where $\frac{lg-dfs}{dfs}$ is near to zero, which are meaningless in practice. So we set the threshold value as 0.001, which means that both two algorithms will be considered to perform the same in the token pairs if $\frac{lg-dfs}{dfs}\leq 0.001$. Then we calculate the percent of token pairs with $\frac{lg-dfs}{dfs}> 0.001$ for different $M$s.

\begin{figure}[H]
    \centering
    \includegraphics[width=1\linewidth]{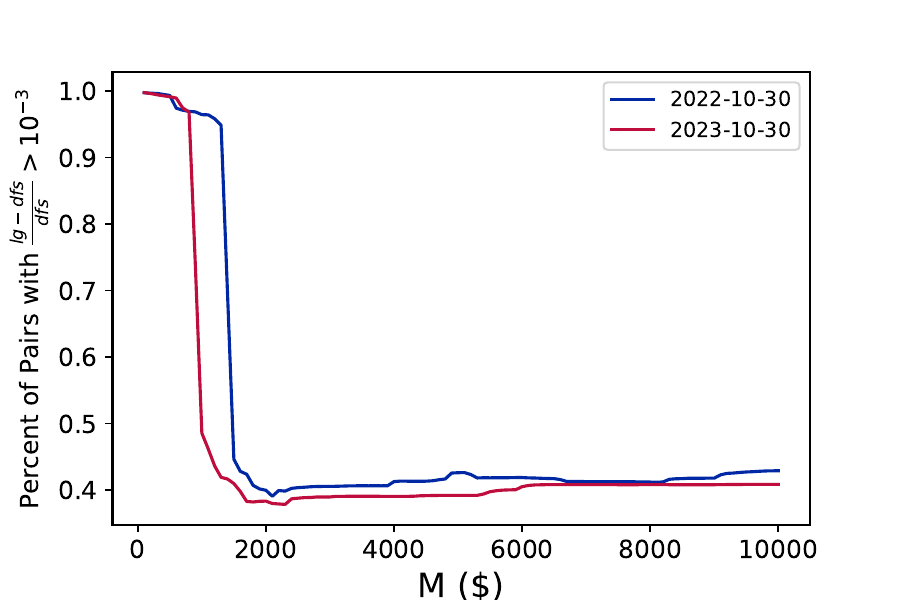}
    \caption{Proportion of Token Pairs with $\frac{lg-dfs}{dfs}> 0.001$ under Different $M$. $M$ denotes the money input in trading for each pair of tokens.}
    \label{nonzero}
\end{figure}

The result is shown in Fig.\ref{nonzero}. When $M<1,000\$$, almost all pairs' $\frac{lg-dfs}{dfs}$ is larger than 0.001, while the figure is around 40\% when $M>1,000\$$.

\begin{figure}[H]
    \centering
    \includegraphics[width=1\linewidth]{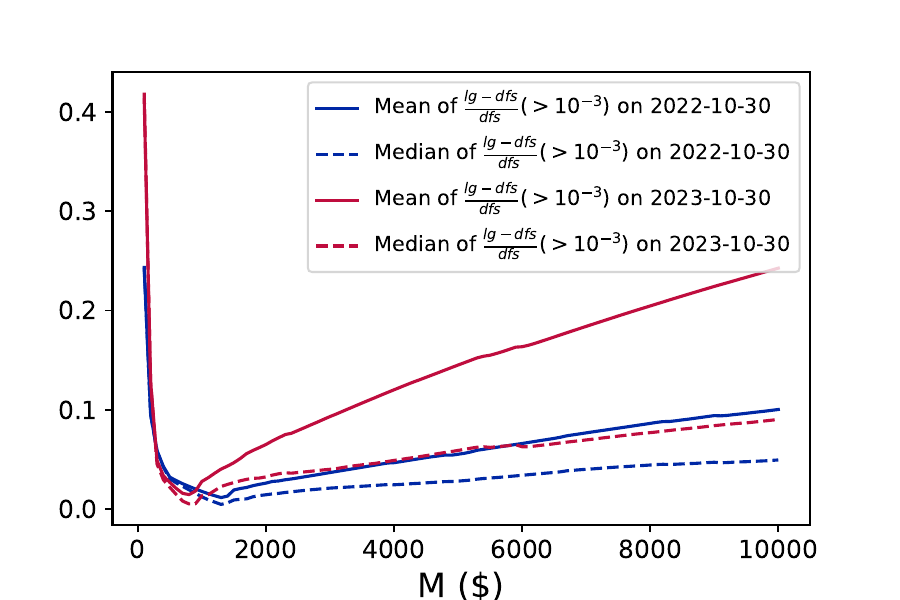}
    \caption{Mean and Median of $\frac{lg-dfs}{dfs}$ with $\frac{lg-dfs}{dfs}> 0.001$ under Different $M$. $M$ denotes the money input in trading for each pair of tokens.}
    \label{meanmedian}
\end{figure}

For all those token pairs where $\frac{lg-dfs}{dfs}> 0.001$, we also calculate their mean and median value with different $M$, which is shown in Fig.\ref{meanmedian}. By focusing on the mean and median in conditions of different $M$, we can find that: when $M<1,000\$$, the values of token pairs' $\frac{lg-dfs}{dfs}$ are centric and decrease with larger $M$; while they are more sparse and increase with larger $M$ when $M>1,000\$$.

In the above statistical analysis, we didn't take gas fees into consideration which is measured in Ether. If we need to contain the gas fee or set some conditions for the gas fee, then some parts of the algorithm are necessary to change, which is not a hard task. However, the settings about the gas fee cost are a bit subjective, which is why we didn't consider it in our algorithm. 

However, we can still reflect the gas fee cost of the two algorithms by comparing the corresponding optimal routing path length because the shorter a routing path is, the less the gas fee. 

\begin{figure}[H]
\begin{subfigure}{0.5\textwidth}
\centering
\includegraphics[width = 1\linewidth]{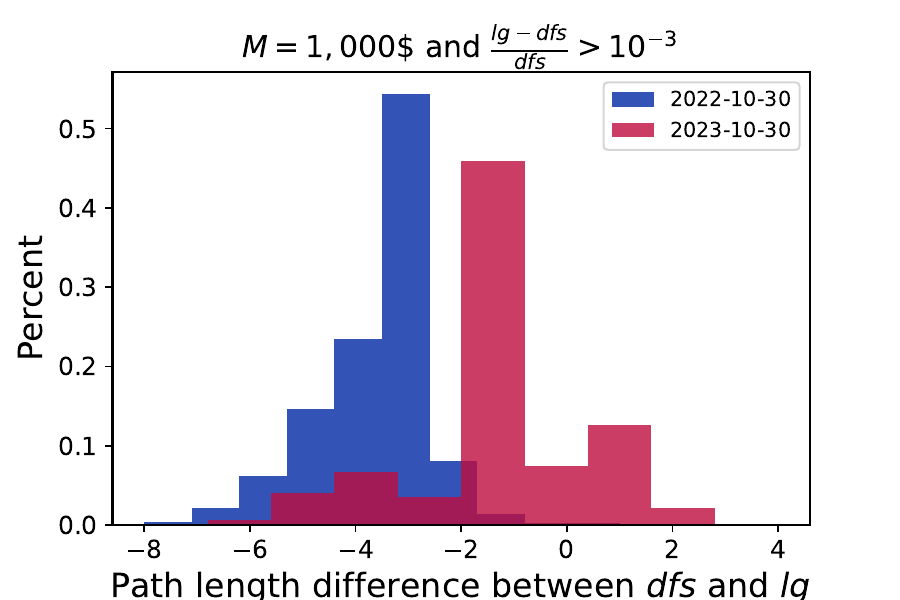}
\caption{$M = 1,000\$$}
\label{path_a}
\end{subfigure}  
\begin{subfigure}{0.5\textwidth}
\centering
\includegraphics[width = 1\linewidth]{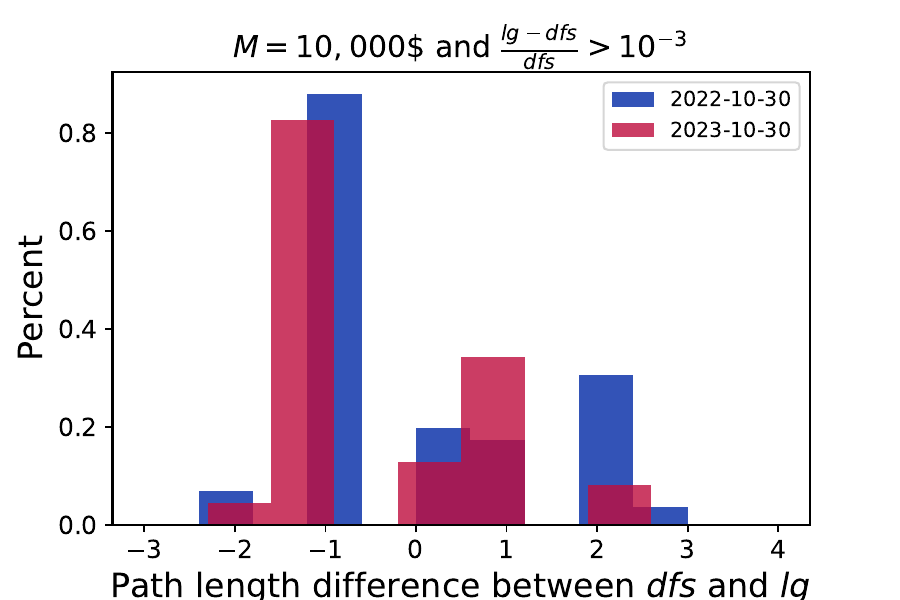}
\caption{$M = 10,000\$$}
\label{path_b}
\end{subfigure}   
\caption{Routing path length comparison of line-graph-based algorithm ($lg$) and DFS algorithm ($dfs$). For any specific pairs of tokens, if the routing path length using $dfs$ algorithm is $l_{dfs}$ and the corresponding path length using $lg$ is $l_{lg}$, then the path length difference between $dfs$ and $lg$ equals $l_{dfs}-l_{lg}$. }
\label{path_length_comp}
\end{figure}

We compare the length of optimal routing paths of the two algorithms in the setting of $M=1,000\$$ and $M=10,000\$$ for pairs where $\frac{lg-dfs}{dfs}> 0.001$. The results are shown in Fig.\ref{path_length_comp}. 

By the case of $M=1,000\$$ (Fig.\ref{path_a}) and $M=10,000\$$ (Fig.\ref{path_b}), we can find that the two algorithms cost almost comparable gas fees in almost all token pairs where $\frac{lg-dfs}{dfs}> 0.001$. Because the line-graph-based algorithm is usually much more profitable than the DFS algorithm, it means that the line-graph-based algorithm outperforms the DFS algorithm comprehensively. 
The results of these comparisons also show the large potential of the line-graph-based algorithm in the practical application of DEXs' routing.

\section{Discussion}

\begin{figure}[H]
\centering
\includegraphics[width = 0.7\linewidth]{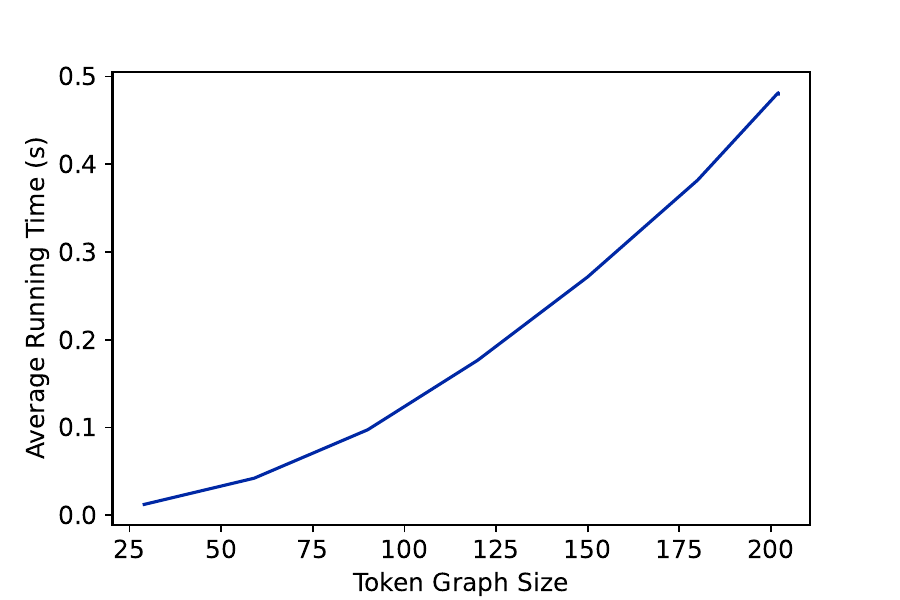}
\caption{Running time (second) of line-graph based method (LG) in token graphs with different sizes}
\label{running_cost}
\end{figure}

In this paper, we have provided a new competitive line-graph-based method to detect an optimal routing path for trading in a DEX, like Uniswap V2, after testing against the DFS method. But how is its performance in running costs? As shown in Fig. \ref{running_cost}, its average running time almost increases exponentially, which may limit its application in large graphs. However, it is still very fast. For example, for a graph with 100 tokens, the average running time is about 0.1s. While the time is around 1.5s if we use the convex optimization method. 
The LG algorithm extends beyond single DEXs to DEX aggregators, to be explored in future work. However, its higher computational complexity compared to DFS remains a limitation, which we aim to address in future optimization research.

\bibliographystyle{IEEEtran}
% \bibliography{reference.bib}

\nocite{*}

\section{Appendix}\label{appendix}

\subsection{Constant AMM in Uniswap V2}\label{amm}
The constant AMM equation in any liquidity pool of Uniswap V2 takes the form of
\begin{displaymath}
        [x+(1-\lambda) \Delta x](y-\Delta y) = xy=k,
\end{displaymath}
where $x$ and $y$ are constant and denote the reserve of token $X$ and $Y$ before trading in the liquidity pool, respectively. $k$ equals the product of $x$ and $y$, $\lambda$ is the transaction tax rate. $\Delta x$ and $\Delta y$ denote the input number of token $X$ and the output number of token $Y$ in trading. By simple derivation, we can get the following function:
\begin{displaymath}
    \Delta y= y-\frac{x\cdot y}{x+(1-\lambda)\cdot \Delta x}.
\end{displaymath}
This equation quantifies the number of token $Y$ ($\Delta y$) that traders can get by entering a specific number of token $X$ ($\Delta x$) in trading.

\subsection{Filtering Token Liquidity Pools}\label{poolfilter}

For a specific considered date (30 October 2023 and 30 Octorber, 2022), we take the following steps to filter tokens and pools:
\begin{itemize}
    \item Select pools whose created date is before the specific date and where the transaction still occurs after the specific date to ensure that the selected pools are still active.

    \item Select pools with more than ten thousand dollars (\$) locked total values (TVL), which can ensure that the liquidity of the pools is deep enough. Traders will incur reduced price slippage when pools are deeper.

    \item Build the token graph with nodes representing tokens and edges representing liquidity pools. Delete nodes whose degree is less than two iteratively.

    \item We set the number of nodes in the token graph as one hundred and start by deleting nodes with the smallest TVL iteratively. The deleting rules are as follows: (1) delete the nodes with the smallest TVL; (2) delete nodes whose degrees are less than 2; (3) if there are still more than one hundred nodes left in the token graph, repeat step (1) and (2) until the number of nodes is equal to or less than one hundred.
\end{itemize}

\subsection{DFS Linear Routing Algorithm}\label{appendix_dfs}
The DFS linear routing algorithm employed by Uniswap V2 to identify optimal trading routes is detailed in Uniswap's smart-order-router repository\footnote{https://github.com/Uniswap/smart-order-router/tree/main/src/routers/}. For clarity, we present its pseudocode below.
\begin{algorithm}[H]
\caption{DFS Linear Routing Algorithm}
\label{Algorithm 6}
    \hspace*{\algorithmicindent} \textbf{Input}: token graph $G$; $\epsilon_1$ units of token $v_1$\\
    \hspace*{\algorithmicindent} \textbf{Output}: The maximal units of token $v_n$
    \begin{algorithmic}
    
    \State marked $\gets$ queue
    \State Path $\gets$ a set to store all detected path from token $v_1$ to token $v_n$

    \Function{dfs}{token graph: $G$, token: $v_1$}
    
        \State marked[$v_1$] = True
        \For{$w$ $\in$ Neighbour($s$)}

            \If{$w=v_n$}
                Path.add(the path to $w$)
            \EndIf
            
            \If{!marked[$w$]}
                DFS($G$, $w$)
            \EndIf
        \EndFor
    \EndFunction
    
    \State $O_{path}$ $\gets$ optimal path.
    \State $O_{v_n}=0$ $\gets$ maximal quantity of $v_n$.
    \State $F(\epsilon_1, p)$$\gets$ A function to calculate the quantity of token $v_n$ by inputting $\epsilon_1$ units of $v_1$ under the path $p$.

    \For{p $\in$ Path}
        \State $out = F(\epsilon_1, p)$
        \If{$out>O_{v_n}$}
        
            \State $O_{v_n}=out$
            \State $O_{path}=p$
        \EndIf
    \EndFor
    
    \Return $O_{path}$, $O_{v_n}$
    \end{algorithmic}
\end{algorithm}

\subsection{Routing Example by Convex Optimization}\label{problem_convex_optimization}

\begin{figure}[H]
\centering
\includegraphics[width = 1\linewidth]{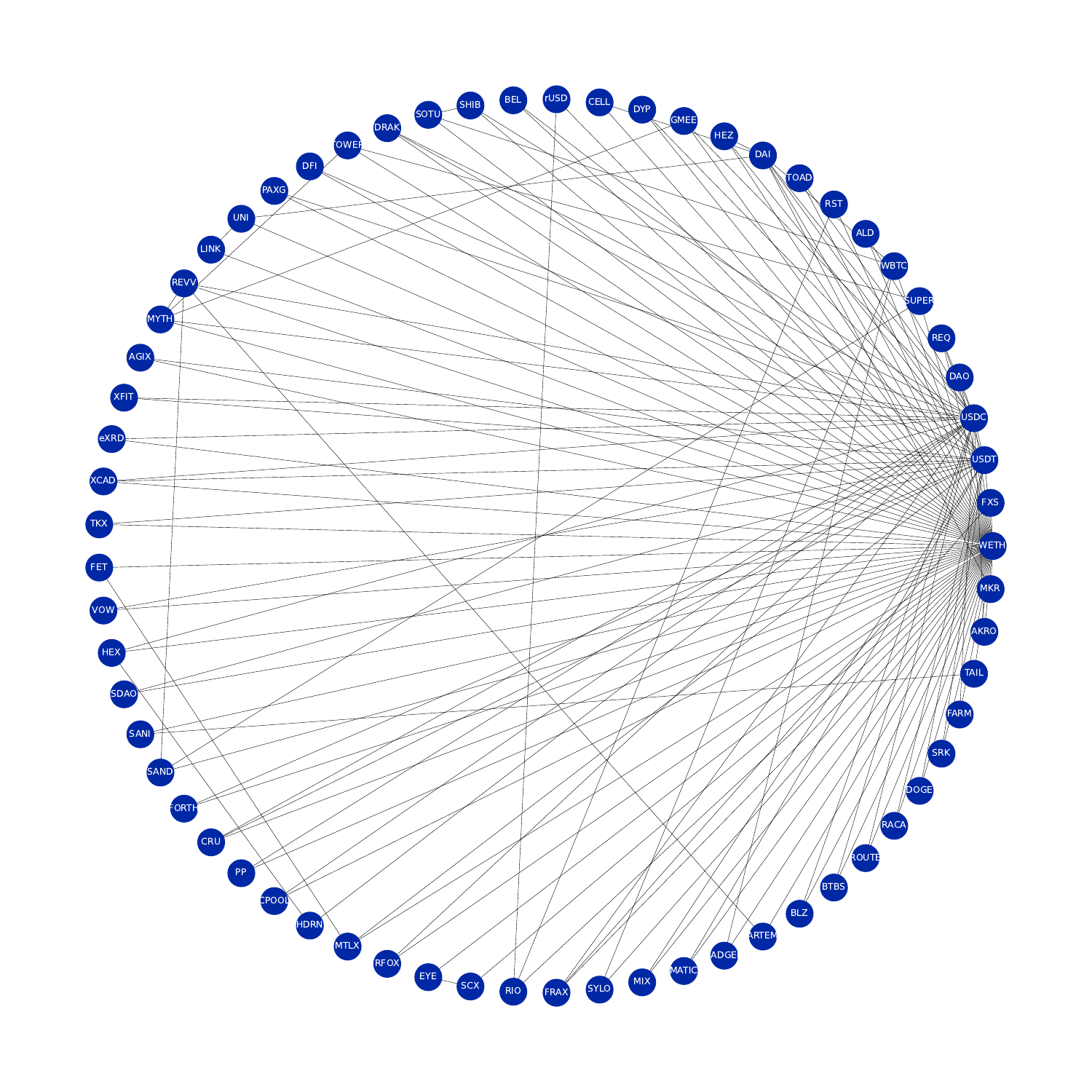}
\caption{Token graphs with 65 tokens and 134 liquidity pools with data from Uniswap V2 on 2023-10-30.}
\label{token_graph2023}
\end{figure}

We built a token network with liquidity pool data from Uniswap V2 on 2023-10-30. By filtering data described in \ref{poolfilter}, we obtain a token graph with 65 tokens and 134 liquidity pools, as in Fig.\ref{token_graph2023}, which includes tokens $WBTC$ and $USDT$. We made a trade experiment by selling 0.01 $WBTC$ to buy $USDT$ with convex optimization. The convex optimization model we used is the same as that in \cite{angeris2022optimal}. For the calculation of the convex optimization method, we use the CVXPY package, and the original code is from the GitHub project "https://github.com/angeris/cfmm-routing-code/blob/master/arbitrage.py". 

\begin{figure}[H]
\centering
\includegraphics[width = 1\linewidth]{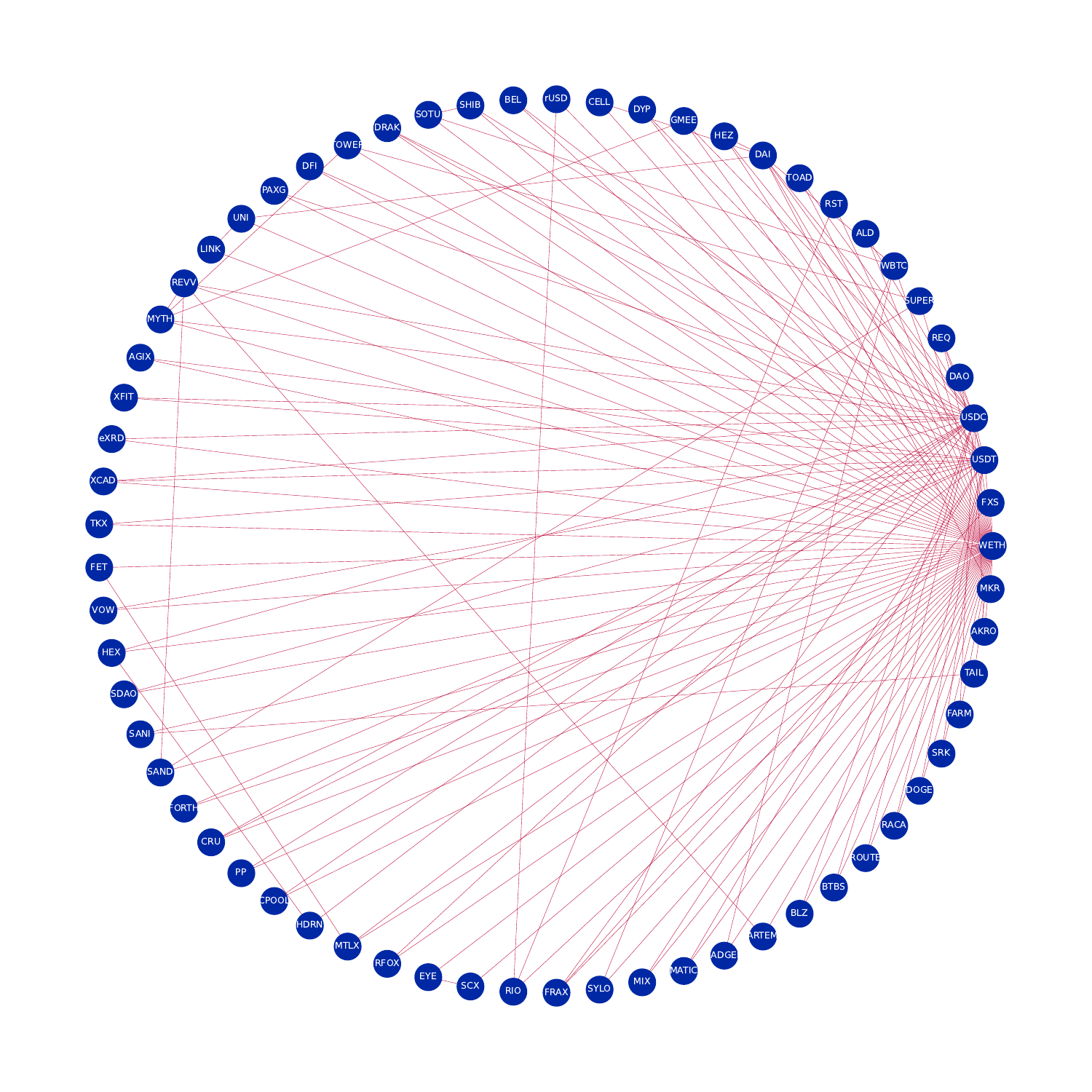}
\caption{Routing results with convex optimization method by selling 0.01 $WBTC$ to by $USDT$.}
\label{token_graph2023_convexoptimization}
\end{figure}

Our trade task is solved successfully according to the status of the convex optimization task with CVXPY (prob.status, where prob is our convex optimization task), and each constraint is satisfied according to the indicator cp.transforms.indicator($cons$), where $cons$ are constraints in the convex optimization task. Then, we check the variables' values that correspond to tokens' inputs and outputs in the convex optimization task. If any tokens' inputs or outputs in a liquidity pool are non-zero, then it means that the corresponding liquidity pool will be part of the routing in our trading experiment, and the color of the corresponding edge (pool) will be changed to red. The result shows that all liquidity pools need to be used in this trade experiment after calculating using the convex optimization method, as shown in Fig.\ref{token_graph2023_convexoptimization}

\textbf{Claim}

The intellectual property rights of this paper are reserved by the authors. Any commercial application or development requires their approval.

% \nocite{*}

\end{document}